\documentclass{SciPost}

\hypersetup{
    colorlinks,
    linkcolor={red!50!black},
    citecolor={blue!50!black},
    urlcolor={blue!80!black}
}

\IfFileExists{mathdesign.sty}{\usepackage[bitstream-charter]{mathdesign}}{\usepackage{amssymb}}
\DeclareSymbolFont{usualmathcal}{OMS}{cmsy}{m}{n}
\DeclareSymbolFontAlphabet{\mathcal}{usualmathcal}

\usepackage{amsthm,mathtools,bm,booktabs,tabularx,enumitem}

\newtheorem{theorem}{Theorem}

\theoremstyle{remark}

\newcommand{\Dcal}{\mathcal{D}}
\newcommand{\Bcal}{\mathcal{B}}
\IfFileExists{dsfont.sty}{\usepackage{dsfont}\newcommand{\idm}{\mathds{1}}}{\newcommand{\idm}{\mathbf{1}}}
\newcommand{\Ccal}{\mathcal{C}}
\newcommand{\Mcal}{\mathcal{M}}

\newcommand{\Acal}{\mathcal{A}}

\fancypagestyle{SPstyle}{
\fancyhf{}
\lhead{\colorbox{scipostblue}{\bf \color{white} ~SciPost Physics }}
\rhead{{\bf \color{scipostdeepblue} ~Submission }}

\fancyfoot[C]{\textbf{\thepage}}
}

\begin{document}

\pagestyle{SPstyle}

\begin{center}{\Large \textbf{\color{scipostdeepblue}{
Thermodynamic cost of locally maintaining a steady state
}}}\end{center}

\begin{center}\textbf{
Sarang Gopalakrishnan\textsuperscript{1$\star$}
}\end{center}

\begin{center}
Electrical and Computer Engineering, Princeton University, Princeton, NJ 08544, USA
\\[\baselineskip]
$\star$ \href{mailto:sgopalakrishnan@princeton.edu}{\small sgopalakrishnan@princeton.edu}
\end{center}

\section*{\color{scipostdeepblue}{Abstract}}
\textbf{\boldmath{%
The Gibbs states of local Hamiltonians have short-range conditional
correlations.  To maintain a steady state with long-range conditional
correlations, one might expect
that a stochastic process must ``pay'' either in nonlocal moves or
entropy production.  Surprisingly, we find strictly local Gaussian
processes that obey detailed balance and have unique steady states
with arbitrarily long-range conditional correlations.  We show that such processes
must still produce entropy, if implemented using a local quadratic
Hamiltonian coupled locally to equilibrium, rapidly relaxing baths.  We
lower-bound this entropy production in terms of the steady state's
conditional mutual information (CMI); if the CMI decays exponentially with distance, the associated length-scale is called the Markov length $\xi$.  The
entropy production per site per relaxation time obeys
$\hat\sigma \ge c \xi^2$ for large $\xi$, and with $c$ an implementation-dependent constant.
We also bound the tradeoff between coupling range and entropy
production.  
We contrast these Gaussian processes with jump processes on discrete state spaces, for which, under certain irreducibility assumptions, steady states with long-range CMI are only possible if detailed balance is violated. We conjecture (and present evidence) that rapidly mixing local jump processes cannot stabilize steady states with long-range CMI. 
}}

\vspace{\baselineskip}

\noindent\textcolor{white!90!black}{%
\fbox{\parbox{0.975\linewidth}{%
\textcolor{white!40!black}{\begin{tabular}{@{}lr@{}}%
  \begin{minipage}{0.57\textwidth}%
    {\small Copyright attribution to authors. \newline
    This work is a submission to SciPost Physics. \newline
    License information to appear upon publication. \newline
    Publication information to appear upon publication.}
  \end{minipage} & \begin{minipage}{0.37\textwidth}
    {\small Received Date \newline Accepted Date \newline Published Date}%
  \end{minipage}
\end{tabular}}
}}
}

\setlength{\emergencystretch}{2em}

\vspace{10pt}
\noindent\rule{\textwidth}{1pt}
\tableofcontents
\noindent\rule{\textwidth}{1pt}
\vspace{10pt}

\section{Introduction}
\label{sec:intro}

Gibbs states of local Hamiltonians are unusual states, and the
canonical processes that stabilize them --- Glauber dynamics and its
variants --- are unusual processes.  In both cases what is unusual can
be stated precisely.  By the Hammersley--Clifford theorem, local Gibbs
states are precisely the full-support distributions with the Markov
property: if we divide the system into regions $A, B, C$ such that the
Hamiltonian decomposes as $H_{AB} + H_{BC}$, then conditioning on the
state of $B$ removes all correlations between $A$ and $C$ --- the
conditional mutual information (CMI) $I(A{:}C\mid B)$ vanishes.  
The analogous constraint for processes like Glauber dynamics is detailed balance with respect to the target steady state.  
Detailed-balance processes model
the interaction of the system with a heat bath; in the steady state
the system has reached equilibrium with the bath, and no entropy is
produced.  One might expect that any local process producing no
entropy in its steady state is of this Glauber type --- in detailed
balance with respect to some local Hamiltonian --- and hence has a
Markov steady state.  Processes that violate detailed balance
generate entropy, quantifiable in terms of the violation
\cite{Seifert2012}.  In many settings, however, the process is a
means of creating a steady state, so the question of most interest
is: how much entropy must be produced to maintain a given steady
state with a local process?

The present work addresses this question, both for Gaussian processes (where explicit calculations are available) and for interacting lattice systems (where our results are more heuristic). 
The answer has some surprising features. First, discrete and continuous state spaces seem fundamentally different. 
One might have expected, on intuitive grounds, that a local process
satisfying detailed balance stabilizes a local Gibbs state. 
While true for discrete jump processes with irreducible frozen blocks and full-support steady states, this is not true for general Gaussian processes: we present
an explicit counterexample --- a hyperuniform steady state with
exactly Coulomb conditional correlations.  
Even though the counterexample
satisfies detailed balance --- so its entropy production is invisible
within stochastic thermodynamics --- we show that any implementation
of it as a \emph{complete, locally interacting} Gaussian system with bounded quadratic stiffness and finite-range linear couplings to heat
baths must nevertheless produce entropy, and we lower-bound that
entropy production in terms of the steady state's long-range
CMI.\footnote{Here and elsewhere we mean the ``true'' CMI,
conditioning on \emph{all} the information in region $B$.  One can
manufacture long-range CMI, even in equilibrium, by hiding or
marginalizing over some of this information.}  Suppose
the CMI decays at long distances with a characteristic \emph{Markov
length} $\xi$: $I(A{:}C\mid B) \sim \iota\,e^{-|B|/\xi}$ with $O(1)$
amplitude $\iota$.  For Gaussian field theories coupled to multiple baths at different temperatures we show
that, in cutoff units (we set the ultraviolet cutoff $\Lambda = 1$
throughout), $\xi \lesssim r\sqrt{T_{\rm hottest}/T_{\rm coldest}}$,
where $r$ is the coupling range.  For our regulated counterexample at
fixed range the entropy production is at least $c_r\xi^2$ per site per
characteristic relaxation time, and we give an explicit nearest-neighbor
implementation attaining this scaling.\footnote{The
time on which small deviations from the steady state relax; this can
be much shorter than the mixing time, and in interacting settings the
distinction can matter (Sec.~\ref{sec:discussion}).} In particular, the algebraic-CMI endpoint of this family requires infinite entropy production within this implementation class. 
That these processes are out of equilibrium is witnessed by a violation of the fluctuation-dissipation theorem: if they are local, their response to perturbations of the Hamiltonian necessarily reveals a mode-dependent temperature. 

For lattice models with a finite local state space, the situation is strikingly different: assuming irreducibility of frozen blocks and full support, one can show that any process with a long-range CMI steady state must violate detailed balance; one can, moreover, quantify this violation (and thus the entropy production) under a uniform entropy-relaxation assumption for the frozen blocks. 
Whether this violation is achievable, we have not been able to prove. We conjecture that for strictly classical processes it is not: shot noise appears to fundamentally limit the achievable Markov length. We show, moreover, that for binary finite-state hidden Markov models with a strictly positive hidden transition matrix and an active odd-flip rule, a long-range CMI is not achievable; this suggests, but does not establish, that rapid-mixing processes cannot reach steady states with long-range CMI. 
Adapting a very recent result of Ref.~\cite{NegariZouHsieh2026}, however, we observe that quantum spin systems with a discrete state space can violate this constraint, and, with a local open-boundary completion, maintain unique rapidly mixing steady states with long-range CMI. The thermal implementation we give has an infinite entropy production rate at this endpoint. 

The rest of this work is structured as follows. Sec.~\ref{sec:background} briefly reviews entropy production and CMI. Sec.~\ref{sec:gaussian} addresses linear stochastic field theories, for which all relevant quantities are Gaussian integrals. Sec.~\ref{sec:jump} turns to local jump processes on discrete state spaces, focusing on processes that relax rapidly to their unique steady state. Finally, Sec.~\ref{sec:discussion} summarizes our results and lists open questions.

\section{Context}
\label{sec:background}

\subsection{What entropy production means}

In the fundamental picture, the system is in contact with several
thermal baths at distinct temperatures.  In steady state the system
itself is unchanged, but energy flows through
it, from hot baths to cold ones.  For two baths at temperatures
$T_1 > T_2$ coupled to a single overdamped degree of freedom in a
unit-stiffness trap, with mobilities
$\Mcal_1, \Mcal_2$, the steady heat current is
$J = \Mcal_1\Mcal_2(T_1 - T_2)/(\Mcal_1 + \Mcal_2)$ [we will define
this carefully for many-body systems later], and the entropy
production~\cite{Godreche2019} is the Clausius rate\footnote{Note that this rate
diverges when the cold bath is at zero temperature.  This is just
$\partial S_{\rm bath}/\partial E = 1/T \to \infty$: a cold bath
gains unbounded entropy per unit heat, so unless the rate of energy
transfer into it is throttled the entropy production rate is
infinite.}
\begin{equation}
  \sigma \;=\; J\Big(\frac{1}{T_2} - \frac{1}{T_1}\Big)
  \;=\; \frac{\Mcal_1\Mcal_2\,(T_1 - T_2)^2}
             {(\Mcal_1 + \Mcal_2)\,T_1 T_2} \;\ge\; 0 .
  \label{eq:clausius}
\end{equation}
This physical entropy production depends on the details of how the
system is coupled to its baths.  What we normally observe is the
\emph{marginal} process acting on the system alone, and this does
not fix the implementation uniquely.  Without further
qualifications, therefore, ``the entropy production needed to
maintain a given state'' is ill-defined.
At one extreme, coupling the system to a mixture of two otherwise
identical heat baths at different temperatures can give rise to a
marginal process that is detailed balanced at some intermediate
temperature: thus, a given physical implementation can produce
``superfluous'' entropy that can be eliminated without changing
either the steady state or any of the observed rates.
At the other extreme, any full-rank state is $e^{-H}$ for some $H$,
and Glauber dynamics with respect to that $H$ maintains it with a
single bath at zero entropy production.
When we talk about the entropy production needed to maintain a
steady state we really mean a constrained infimum over all
\emph{local} implementations that maintain $e^{-H}$ as the system's
steady state.

The standard entropy production in stochastic thermodynamics is a
different constrained infimum: over all implementations (local or
not) that reproduce \emph{each} rate of the marginal
process.  This infimum
is~\cite{Schnakenberg1976,Seifert2012,EspositoVanDenBroeck2010,GomezMarin2008}
\begin{equation}
  \sigma \;=\; \sum_{x,u} \pi(x)\,W_u(x)\,
  \ln\frac{\pi(x)\,W_u(x)}{\pi(x+u)\,W_{-u}(x+u)},
  \label{eq:sigmainf}
\end{equation}
and is attained by assigning each inverse pair of transitions to a
separate bath, whereupon the bath temperature and exchanged heat
enter only through the combination
$q/T = \ln[W_u(x)/W_{-u}(x+u)]$ (splitting a pair across several
baths only adds to the total, by the log-sum inequality).
The concept we focus on and the standard one are similar in both
being infima, but under inequivalent constraints.  A given marginal
process that is promised to have a local implementation must satisfy
both constraints, and must therefore produce at least as much
entropy as the more stringent of these bounds.

\subsection{Conditional mutual information}

Almost everywhere in this work we will be concerned with the classical CMI, $I(A:C|B)$, which is simply the average mutual information of the conditional distribution: $\sum_b p_B(b)\, I(A{:}C)_{P(AC|b)}$. In terms of Shannon entropies one can write $I(A:C|B) = S_{AB} + S_{BC} - S_B - S_{ABC}$, where subscripts denote subsystems. This latter definition carries over to quantum systems if one replaces Shannon entropies with von Neumann entropies. Our main partitioning geometry is the annular one shown in Fig.~\ref{fig:cmi}(a), where $B$ separates $A$ and $C$ by some distance $|B|$ at its narrowest. For a full-support classical distribution on a finite graph, vanishing CMI for every separating tripartition is equivalent to a Gibbs representation by interactions on the graph cliques---this is the Hammersley-Clifford theorem. In particular, a range-$r$ Hamiltonian gives zero CMI for every complete shielding partition wider than $r$. Hammersley-Clifford has not been fully generalized yet for quantum Hamiltonians, but see Refs.~\cite{HaydenJozsaPetzWinter2004, KKB2020, chen2025quantum, bakshi2026dobrushin}. In both classical and quantum settings, a small CMI implies efficient local reconstruction: specifically, if $I(A:C|B) \leq e^{-|B|/\xi}$ for fixed $\xi$, one can approximately retrieve $P_{ABC}$ after $A$ is erased, by a recovery channel acting on $B$, with error controlled by the CMI~\cite{HaydenJozsaPetzWinter2004, fawzi2015quantum, junge2018universal}. Thus the CMI plays a central part in the theory of quantum error correction, mixed-state phases of matter \cite{SangHsieh2025, 1kr5-jlcb}, and learnability transitions \cite{KumarEtAl2026, Hauser2026}---including retrieval in neural-network diffusion models~\cite{hu2025local}. The ubiquity of the CMI in these settings makes it natural to ask how one might control it, and what resources such control would require.

An important distinction between CMI and other standard information-theoretic quantities is that CMI does not necessarily decrease under marginalizing: a Markov chain acting on $B$ can either increase or decrease $I(A:C|B)$~\cite{PRXQuantum.4.040332, PhysRevA.110.032426, PhysRevLett.133.200402, SangHsieh2025,ZhangGopalakrishnan2025}. In particular, marginalizing a Gibbs state will generically leave it non-Gibbs~\cite{vEFS1993}: for example, the critical Gibbs states of a short-range Ising Hamiltonian has short-range CMI but long-range MI, so a Markov chain that simply deletes $B$ will replace the short-range CMI with the long-range MI. This is why it will be crucial in the rest of this work to focus on the case where one conditions \emph{completely} on the state of $B$.

\subsection{What our results could not be}
\label{sec:context}

To orient the rest of the discussion, we note a few conjectures one
might have been tempted to make but that are ruled out by elementary
considerations or existing results.

\begin{enumerate}
\item Entropy production is a rate and the Markov length is not, so
any relation between them involves entropy produced per timescale.
For steady states the natural timescale is the one on which local
deviations from the steady state (of the kind the process continually
generates) relax, not the global mixing time: the stationary entropy budget concerns the continual
relaxation of fluctuations, rather than preparation from an arbitrary
initial state.  Because the relaxation time diverges at an equilibrium
critical point, our bounds constrain critical processes only weakly;
indeed, arbitrarily weak detailed-balance-violating perturbations of
a critical point --- producing arbitrarily little entropy ---
generate long-range CMI, as shown very recently in
Ref.~\cite{ChenGrover2026}.  
\item Our results assume a unique steady state for the specified finite-volume dynamics; uniformity over volumes and frozen boundary conditions is stated separately.  Dynamics with multiple steady states ---
conserving model-B dynamics, say --- generate long-range steady-state
CMI for free: under high-temperature charge-conserving equilibrium
dynamics, any initial state of definite charge evolves to a
distribution that is random apart from the global charge constraint,
which suffices for long-range CMI, with no entropy produced. 
\item Our results concern steady states, not transients.  In the
recent literature on finite-time phase transitions --- e.g.,
spontaneous strong-to-weak symmetry breaking of product states under
stochastic dynamics (see \cite{Wang2026SWSSB} for a recent review) ---
the CMI diverges \emph{at} the finite-time transition, but not
in the steady state of the associated process.
\item For discrete-time Markov chains it is straightforward to write
down ``prepare-and-replace'' dynamics with long-range CMI in the
steady state\footnote{Carolyn Zhang, private communication.}: run stochastic dynamics from a fixed configuration up
to the time at which the CMI diverges (as above), reset, and
alternate indefinitely.  The entropy accounting of such discrete-time
protocols is deferred to future work.
\item Any bound must rely, not just on the Markov length, but also on the amount of CMI that is present at large distances. An illustrative example was very recently studied in Ref.~\cite{ChenGrover2026}, consisting of two non-reciprocally coupled Gaussian fields near criticality. Infinitesimal coupling suffices to mix the correlation and Markov lengths, but can clearly only produce entropy at an infinitesimal rate. 
\item There can be no general converse, from detailed-balance
violation to non-Gibbsian steady states: many processes violate
detailed balance yet mix to the infinite-temperature state --- for
example, independent three-state spins whose only allowed transitions
are $0 \to 1 \to 2 \to 0$, at equal rates.
\end{enumerate}

\subsection{Relation to previous work}

The question addressed here---minimal entropy production needed to maintain a target state---was previously explored in Ref.~\cite{HorowitzEngland2017}; our setup differs from this work, and other previous literature on entropy production, by explicitly imposing a spatial restriction on the allowed processes. The field theories we consider are closely related to those explored in Refs.~\cite{Nardini2017, LiCates2021}; in particular, their ``model AB'' exhibits the phenomenon of a local process satisfying detailed balance with respect to a nonlocal parent Hamiltonian. 
The distinction between resolved and hidden entropy production is
also central to coarse-grained stochastic thermodynamics
\cite{Esposito2012,GomezMarin2008}.
The steady state of the counterexample belongs to a family of states that have been explored in the context of hyperuniformity~\cite{HexnerLevine2017, Torquato2018, MaireChaix2025, Casiulis2026}.
The Markov property for quantum systems has seen considerable recent attention~\cite{chen2025quantum, BakshiEtAl2025, RosaRuizEtAl2026}, though primarily for dynamics that is explicitly detailed balance with respect to a local parent Hamiltonian.

Finally, two very recent works (posted while this paper was nearing completion) address similar themes: Ref.~\cite{ChenGrover2026} showed that the steady-state Markov length can diverge for equilibrium critical points subject to weak nonequilibrium perturbations (which automatically produce arbitrarily little entropy), and Ref.~\cite{NegariZouHsieh2026} presented an example of a quantum steady state with long-range Markov length, which offers an interesting contrast to the classical discrete setting, where these states seem not to exist.

\section{Single-component Gaussian processes}
\label{sec:gaussian}

\subsection{Setup}
\label{sec:setup}

In this section we consider statistically translation-invariant Ornstein--Uhlenbeck theories of the form
\begin{equation}
  \dot\phi = -\Bcal\phi + \eta, \qquad
  \langle \eta(\mathbf x,t)\eta^T(\mathbf x',t')\rangle
   = 2\Dcal\,\delta(\mathbf x - \mathbf x')\delta(t-t'),
  \label{eq:ou}
\end{equation}
where $\phi$ is an $n$-component real field; $\Bcal(k)$ and
$\Dcal(k)$ are matrix-valued; $\mathcal{B}$ is stable, i.e., has only eigenvalues with positive
real part; and $\Dcal$ is symmetric positive
semidefinite.
Equivalently, the process is specified by its MSR action,
$S = \int dt\, \hat\phi\,(\partial_t + \Bcal)\,\phi - \Dcal\,\hat\phi^2$. 
Real-space locality enforces that $\mathcal{B}(k), \mathcal{D}(k)$ polynomials in $k$, and reality of $\phi$ together with parity symmetry kills odd powers of $k$ in the small-$k$ expansion. 
We define the drift $F[\phi] = -\Bcal\phi$.  Wavenumbers run over the \emph{band}
$|k| \le \Lambda$, with $\Lambda$ the ultraviolet (lattice) cutoff; we
set $\Lambda = 1$. The stationary covariance solves
$\Bcal\Ccal + \Ccal\Bcal^T = 2\Dcal$; the stationary measure is
$\pi \propto e^{-\frac12\phi^TK\phi}$ with $K = \Ccal^{-1}$, and naturally specifies an effective Hamiltonian. We can write 
\begin{equation}
  \Bcal\Ccal = \Dcal + Q, \qquad Q^T = -Q
  \qquad\Longleftrightarrow\qquad \Bcal = (\Dcal + Q)K,
  \label{eq:decomp}
\end{equation}
with $Q$ the \emph{circulation} \cite{Tomita1974,Graham1989}: $Q = 0$ iff
detailed balance holds. 

For Gaussian states, correlations and mutual information
are read off from the covariance $\Ccal$, while conditional correlations
and CMI are read off from its \emph{inverse}, the ``precision matrix'' (or ``effective Hamiltonian'')
$K = \Ccal^{-1}$. Poles of $K$ are zeroes of $\mathcal{C}$ and vice versa. 
Rational $K(k)$ with complex poles gives exponentially decaying CMI, with
a length scale $\xi^{-1} = \min|\mathrm{Im}\,k_{\rm pole}|$; real-axis
singularities give power-law or non-decaying CMI. We use $\xi$ for
this precision-decay scale below; in the one-dimensional setting, the Markov length is $\xi/2$.

\subsection{The counterexample}
\label{sec:reversible}
\label{sec:lattice}
\label{sec:coulomb}

\begin{figure}[t]
\centering
\includegraphics[width=0.66\textwidth]{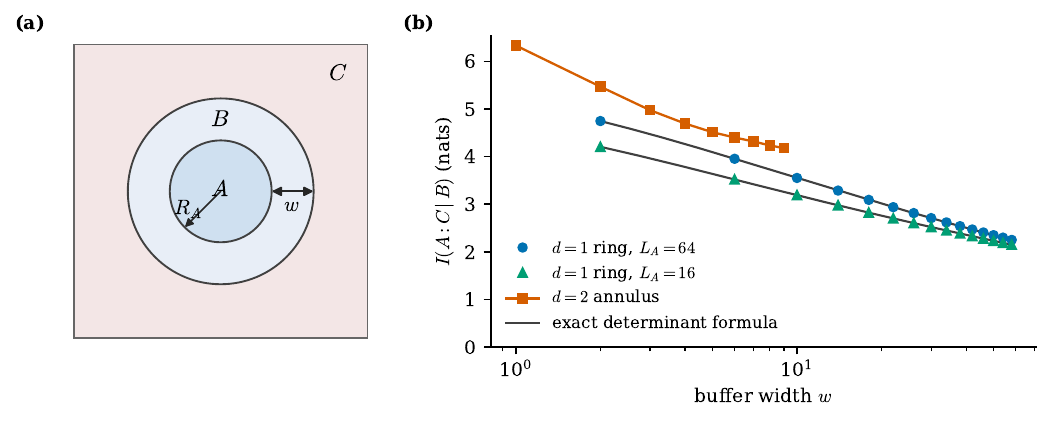}
\caption{(a) The annular tripartition used throughout: a buffer $B$
of width $w$ separates a region $A$ from the rest of the system $C$;
in $d = 1$ the annulus degenerates to the two buffer segments of the
ring geometry used in panel (b).  (b) Exact Gaussian conditional
mutual information of the near-Coulomb state \eqref{counterex},
computed from covariance determinants, with an infinitesimal
regulator ($\Ccal \to \Ccal + T_1$, $T_1 = 10^{-6}$) keeping the CMI
finite.  $d = 1$: ring of $N = 512$ sites, blocks $L_A = 64$ and
$16$, two-sided buffer of width $w$ per side.  Markers: exact
evaluation; lines: the determinant formula
\eqref{eq:cmi2ch} of Appendix~\ref{app:cmi}.
$d = 2$: torus of $40 \times 40$ sites, $A$ a disk of radius $5$, $B$ an
annulus of width $w$.}
\label{fig:cmi}
\end{figure}

Equipped with these definitions, we can introduce the model that
will illustrate many of the themes of this paper: a one-component
field on a chain, with ultralocal drift and conserved
nearest-neighbor noise,
\begin{equation}\label{counterex}
  \dot\phi_i = -\phi_i + \eta_i, \qquad
  \eta_i = \zeta_i - \zeta_{i-1}, \qquad
  \langle\zeta_i\zeta_j\rangle = 2\delta_{ij}
  \quad\Longrightarrow\quad
  \Dcal(k) = 4\sin^2(k/2).
\end{equation}
Since $\Bcal = \idm$, the stationary covariance is
$\Ccal = \Dcal$, and $\Bcal\Ccal = \Dcal$ is symmetric: $Q = 0$, and
detailed balance holds. (The $k = 0$ mode is noiseless but
damped, so it pins to zero, and the steady state is
unique.) All modes relax at the same rate. Correlations
vanish identically beyond one site but
\begin{equation}
  K(k) = \frac{1}{4\sin^2(k/2)}
\end{equation}
has a \emph{real-axis} pole at $k = 0$, so the Markov length is
infinite.  
For consecutive finite intervals $A,B,C$ in the infinite chain, the CMI is (App.~\ref{app:cmi})
\begin{equation}\label{cmidecay}
I(A:C|B) = -\frac{1}{2} \log\left(1 - \frac{|A|}{|A|+|B|+1} \frac{|C|}{|C|+|B|+1}\right),
\end{equation}
so it scales as $\sim 1/|B|^2$ for $|A|, |C|$ fixed and small, $|B|$ large, and $\sim \tfrac12\log(L/|B|)$ for $|A|=|C|=L\gg |B|$. The case of the full system requires care: the state space is continuous, and the CMI naively diverges unless regulated by adding infinitesimal noise.
Fig.~\ref{fig:cmi} plots the regulated ring and the analogous two-dimensional geometry.
%
The
effective Hamiltonian is the confining
one-dimensional lattice Coulomb potential, $K_{ij} \sim -|i-j|/2$ \cite{Hauser2026}.
This example illustrates the counterintuitive fact that a completely local process can satisfy detailed balance with respect to a nonlocal effective Hamiltonian. 

There are two very different ways of reading this process. The first is the ``obvious'' one: $\phi$ is coupled to two baths at different temperatures; the cold bath is at $T = 0$, and obeys no constraints, while the hot bath is at nonzero $T$, and is restricted to charge-conserving moves. (The conserving bath also gives a $k$-dependent damping term, but this is subleading at small $k$ and can be neglected.) 
Also available, though, is an equilibrium reading, according to which the fundamental field is $E_i = \sum_{j\le i}\phi_j$, its Hamiltonian is $H = \tfrac12\sum_i E_i^2$ (nonlocal in the $\phi$ variables), and it undergoes single-site Glauber updates at temperature $T = 1$, $\dot{E}_i = - \partial H / \partial E_i + \zeta_i$, where $\zeta_i$ is white noise. Then $\dot{\phi}_i = \dot{E}_i - \dot{E}_{i-1} = - \phi_i + \zeta_i - \zeta_{i-1}$, precisely Eq.~\eqref{counterex}. 
Thus, the process~\eqref{counterex} itself is equally compatible with strongly nonequilibrium local dynamics and equilibrium dynamics with respect to a nonlocal Hamiltonian. 

These two interpretations of Sec.~\ref{sec:coulomb} describe the same differential equation, so any passive correlation measurement will yield identical answers in the two cases. In particular, $\mathcal{C}(k, \omega) \sim k^2$ as $\omega \to 0$. However, the response to perturbations of the Hamiltonian distinguishes between them. We sketch the main ideas here; a derivation in the Keldysh framework is presented in App.~\ref{app:keldysh}. For a Gaussian theory, the response function $\mathcal{R}(k, \omega)$ of a mode probes its intrinsic dynamics, independent of the occupation factors~\cite{Kamenev_book}. The effective temperature of mode $k$, $T_{\mathrm{eff}}(k, \omega) \equiv \omega \mathcal{C}(k, \omega) / [2\mathrm{Im} \mathcal{R}(k, \omega)]$. 
On the Coulomb interpretation, the low-momentum modes are increasingly stiff, so their response is suppressed as $\mathcal{R} \sim k^2/(\lambda - i \omega)$, similar to their fluctuations: thus, $T_{\mathrm{eff}}$ approaches a constant, as it should since the system is in equilibrium with respect to a nonlocal Hamiltonian. On the nonequilibrium interpretation, the stiffness of the low-$k$ modes is a $k$-independent constant, so $\mathcal{R} \sim 1/(\lambda - i \omega)$ at small $k$. This yields $T_{\mathrm{eff}}(k) \sim k^2$. Thus the fluctuation-dissipation theorem is violated on the nonequilibrium interpretation. Crucially, the Coulomb interpretation requires a singular Hamiltonian stiffness, $\mathcal K_C\sim k^{-2}$. For a local quadratic Hamiltonian with uniformly bounded stiffness $0<\mathcal K(k)\le K_*$, the static response is $\mathcal R(0,k)=1/\mathcal K(k)\ge1/K_*$. Matching the vanishing fluctuations then forces $T_{\mathrm{eff}}(k)=\mathcal K(k)\mathcal C(k)$ to vanish as $k\to0$. Thus any such local interpretation must be nonequilibrium.

\subsection{Generalizations}
\label{sec:generalizations}

Even for a single-component field, the mechanism outlined in the previous section can be extended to give increasingly nonlocal CMI, by forcing the noise to come with higher powers of $k$, e.g., through dipole moment conservation \cite{GromovLucasNandkishore2020}. This mechanism can lead to a ladder of hyperuniform steady states, with $S(k) \sim k^{2n}$ for any integer $n \geq 1$; all these steady states have diverging CMI, and all the associated processes are local and obey detailed balance. More generally, for any rational function $K(k) \equiv \nu(k)/\theta(k)$, one can straightforwardly construct a local process that satisfies detailed balance and has a steady state with precision $K(k)$, as follows. Take $\nu$ and $\theta$ to be even polynomials with $\nu(k) > 0$ on
the band and $\theta(k) \ge 0$ on the band, and choose
\begin{equation}
  \Bcal(k) = \nu(k), \qquad \Dcal(k) = \theta(k).
\end{equation}
Both are polynomials, so the process is local.  Since $\nu > 0$,
every mode is damped: the steady state exists and is unique, with
covariance $\Ccal = \Dcal/\Bcal = \theta/\nu = K^{-1}$, as desired.
And $\Bcal\Ccal = \theta = \Dcal$ is symmetric, so $Q = 0$: detailed
balance holds by construction.  The counterexample is the case
$\nu = 1$, $\theta = 4\sin^2(k/2)$.
The representation is not unique:
multiplying $\nu$ and $\theta$ by a common positive even polynomial
yields a {different} local reversible process with the
{same} steady state.
(The construction generalizes verbatim to $n$-component fields:
$\theta$ the common denominator, $\Dcal = \theta\,\idm$, and
$\Bcal = \theta K$ positive definite on the band.)

In continuum notation the counterexample reads
\begin{equation}
  \dot\phi = -\lambda\phi + \nabla\!\cdot\!\bm\eta, \qquad
  \Dcal = Mk^2, \qquad
  \Ccal(k) = \frac{Mk^2}{\lambda}, \qquad
  K(k) = \frac{\lambda}{Mk^2},
  \label{eq:coulombmodel}
\end{equation}
with $\lambda$ the non-conserving relaxation rate and $M$ the
conserved-noise mobility.  We use this form whenever dimensional
constants matter, as in the bounds of Sec.~\ref{sec:embedding}.


\subsection{Entropy production}
\label{sec:embedding}

We have seen that the counterexample violates the fluctuation-dissipation theorem. We now turn to lower-bounding this violation, and the associated entropy cost, more generally. For simplicity we will continue to focus on the one-component, translation-invariant setting. For the explicit bounds we take $H=\tfrac12\sum_i\phi_i^2$ and finite-range linear couplings to Markovian baths. The first thing we need to establish is that long Markov lengths cannot arise for a local field $\phi$ in equilibrium with a local bath. This is slightly less trivial than it sounds: the composite system obeys the Markov property by assumption, but we are concerned with the marginal over the bath, and the Markov length can grow (indeed diverge) under marginalization. For a particular $k$ mode, after integrating out the bath, the precision matrix for the $\phi$ fields is $K_{\mathrm{eff}}(k) \equiv K_0(k) - h^2(k) G_b(k)$, where $G_b(k)$ is the bath's correlation function at $k$. Stability of the state forces $h^2 \leq K_0 G_b^{-1}$ for every $k$. The Markov length of the system's marginal is set by the pole of $K_{\mathrm{eff}}(k)$ that lies closest to the real axis. By assumption, $K_0, h$ are local, and do not have poles near the real axis. The danger is that $G_b$ might: in other words, integrating out a gapless mode can generate effective long-range interactions. Provided the bath has a finite correlation length $\xi_{\mathrm{bath}}$, however, the nearest pole of $G_b(k)$ is also gapped away from the real axis, so $K_{\mathrm{eff}}(k)$ has no singularities near the axis, and $\phi$ cannot have a large Markov length. To summarize, marginalizing over the bath cannot generate a Markov length that is larger than the bath's correlation length. (Note that a similar result for local state spaces was proved in the high-temperature limit in Ref.~\cite{ZhangGopalakrishnan2025} --- if we take the system and its bath to be in equilibrium at sufficiently high temperature and apply the erasure channel to the bath, Theorem 2 of that work states that the system's marginal retains a finite Markov length.) 

Any remaining, strictly local, route to a large Markov length must come from nonequilibrium effects. To parameterize these we will couple the system to a set of arbitrarily many baths, indexed by $\nu$. Each bath is in equilibrium at a temperature $T_\nu$ and a characteristic, momentum-dependent Golden Rule Rate $\mathcal{M}_\nu(k)$ at which it couples to the system: $\mathcal{M}_\nu(k) \equiv g(k)^2 \nu(k)$, where $g(k)$ is the interaction between system mode $k$ and the bath, and $\nu(k)$ is the bath density of states. Using each bath's FDT, and suppressing mode indices, the marginal process has $\mathcal{B} = \sum\nolimits_\nu \mathcal{M}_\nu, \, \mathcal{D} = \sum\nolimits_\nu \mathcal{M}_\nu T_\nu$. Matching the dissipation $\mathcal{B} = \lambda$, we get that $\mathcal{D} = \lambda T_{\mathrm{eff}}$, where $T_{\mathrm{eff}}(k) = \mathbb{E}_k(T)$ and $\mathbb{E}_k(f) \equiv \lambda^{-1} \sum\nolimits_\nu \mathcal{M}_\nu(k)\, f(T_\nu)$ denotes the average over baths weighted by their couplings to mode $k$. Rearranging the Clausius entropy-production formula gives
\begin{equation}
  \sigma_k = \lambda\big[\,T_{\mathrm{eff}}(k)\,\mathbb E_k[1/T] - 1\,\big],
\end{equation}
which is non-negative for every $k$ by Jensen's inequality. 

In general, achieving a large Markov length requires $\mathcal{D}(k)$ to have a zero near the real axis. Once the range of the process is capped, the only possible origin for this zero is from $T_{\mathrm{eff}}(k)$. Assume for simplicity that the zero is on the imaginary axis, and the behavior around it is quadratic: $\mathcal{D}(k) = \tfrac12\mathcal{D}''(0)(\kappa^2 + k^2)$, $\kappa^2 = 2\mathcal{D}(0)/\mathcal{D}''(0)$. The precision length is $\xi = 1/\kappa$ (with the CMI convention specified in Sec.~\ref{sec:setup}). It is clear that $\lambda T_{\mathrm{coldest}} \leq \mathcal{D}(k) \leq \lambda T_{\mathrm{hottest}}$; moreover, if the coupling range is $r$, $\mathcal{D}(k)$ is a trigonometric polynomial of degree $r$ in $k$. By Bernstein's inequality, a degree-$r$ polynomial that is bounded on the band can have curvature at most $|\mathcal{D}''| \leq r^2 \lambda T_{\mathrm{hottest}}$: a range-$r$ coupling cannot bend faster than $r^2$ times its height. This gives the scaling of the bound. Keeping the full Taylor series and centering between the noise floor and ceiling gives the sharper result
\begin{equation}
  \boxed{\;\xi\le\frac{r}{2\operatorname{arctanh}\sqrt{T_{\rm coldest}/T_{\rm hottest}}}
  \le\frac r2\sqrt{\frac{T_{\rm hottest}}{T_{\rm coldest}}}\;.}
  \label{eq:tempratio}
\end{equation}
The argument above bounds the achievable Markov length in terms of the bath temperatures. 

Next, we fix the range and Markov length, and estimate the minimum entropy production consistent with these. We proceed as follows. Define $g(k) \equiv \mathbb{E}(1/T)$, so the mode-wise entropy production is $\sigma_k = \lambda (g(k) \mathcal{D}(k) - 1)$, with $g(k) \geq 1/\mathcal{D}(k)$. Any choice of bath defines some valid $g(k)$, and for range $r$, $g(k)$ must be an order-$r$ polynomial in $k$, no matter how many baths we couple to the system. To get a bound, we relax the problem and optimize over all polynomials $g(k)$ rather than optimizing over bath assignments: we compute the ``majorant bound''
\begin{equation}
  L_r(m)=
  \min_{\substack{
    g\ \text{real trigonometric polynomial}\\
    \deg g\le r,\;\;g(k)\ge1/D(k)
  }}
  \int_{-\pi}^{\pi}\frac{dk}{2\pi}
  \bigl[D(k)g(k)-1\bigr].
\end{equation}
This lower-bounds the entropy production regardless of the choice of bath. The physical intuition is particularly simple when $m$ is small.
The target
\begin{equation}
  \frac{1}{D(k)}\simeq\frac{1}{m+k^2}
\end{equation}
has a narrow, tall peak: height $1/m$ and width $\sqrt m$.
A polynomial with only $r$ Fourier harmonics cannot follow an
arbitrarily narrow peak. To stay above it, the polynomial must also
be large elsewhere. That unavoidable excess is entropy production.
A helpful way to rewrite this bound is to note that $D(k) g(k) = 0$ at $k = - i \sqrt{m}$. Minimizing $g(k) \mathcal{D}(k) - 1$ subject to these constraints gives the closed-form bound
\begin{equation}\label{eq:exactmajorant}
  \boxed{
    L_r(m)=
    \frac{2}{
      \sinh((r+2)\kappa)/\sinh\kappa-(r+2)
    }.
  }
\end{equation}
For fixed range and large $\xi$,
\begin{equation}
  L_r(m)\sim
  \frac{12\xi^2}{(r+1)(r+2)(r+3)}.
\end{equation}
Thus fixed-range thermal implementations necessarily pay a cost
growing as $\xi^2$. In particular, at nearest-neighbor range,
\begin{equation}
  L_1(m)=\frac{2}{m(m+4)},
\end{equation}
and an explicit two-bath construction attains it.

\begin{figure}[t]
\centering
\includegraphics[width=0.88\textwidth]{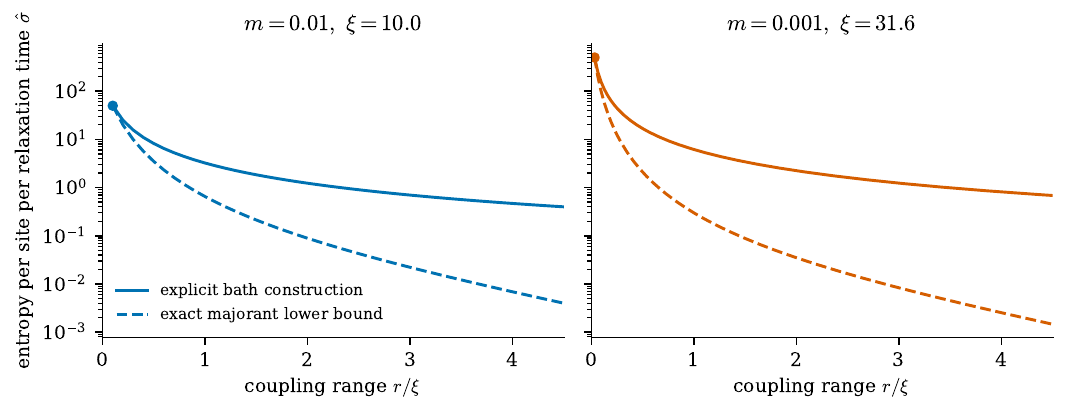}
\caption{The range--dissipation tradeoff for $\Ccal=m+4\sin^2(k/2)$, at $m=0.01$ and $0.001$. Dashed curves: the lower bound from optimizing over \eqref{eq:exactmajorant}. Solid curves: entropy production of the explicit bath construction \eqref{eq:explicitbaths}. The bounds coincide at nearest-neighbor range. Both curves use the lattice momentum measure; the lower bound is not asserted to be physically attainable at higher range.}
\label{fig:dilation}
\end{figure}

We choose the following two baths:
\begin{equation}
 T_c=m,\qquad T_h=m+4,\qquad
 \Mcal_c=\lambda(1-s),\qquad \Mcal_h=\lambda s.
 \label{eq:explicitbaths}
\end{equation}
Their coupling amplitudes are proportional to $(1+e^{ik})/2$ and $(1-e^{ik})/2$, so both are nearest-neighbor couplings. Their entropy production is
\begin{equation}
  \boxed{\;\hat\sigma_k=\frac{16s(1-s)}{m(m+4)},\qquad
  \hat\sigma_{\rm site}=\frac{2}{m(m+4)}\sim\frac{\xi^2}{2}.\;}
  \label{eq:xibound}
\end{equation}
Here $\hat\sigma=\sigma/\lambda$ and $\xi=[2\operatorname{arsinh}(\sqrt m/2)]^{-1}$. At nearest-neighbor range, this choice saturates the bound. 
Figure~\ref{fig:dilation} compares the cost of the explicit implementation with the lower bound from optimization; the bounds coincide at $r=1$. The positive lower bound at every finite $r$ assumes strictly finite-range Markovian baths. As discussed earlier in this section, exponentially decaying CMI can instead arise at equilibrium by marginalizing over a bath with finite correlation length.

\section{Jump processes on discrete spaces}
\label{sec:jump}

In this section we consider jump processes that are defined on discrete systems (e.g., Ising spins or lattice gases): the local state space has a finite number of states, and these are locally coupled by jump moves of bounded support: a configuration $x$ is updated to $x + u$, where $u$ differs from $x$ only locally, and the probability of the jump happening only depends on the local neighborhood of $u$ (within some range $r$). 
We will see that results of the previous section are surprisingly fragile to discreteness. 

\paragraph{Detailed balance + locality implies locality of $H$.}
The counterexample cannot arise for full-support jump processes with irreducible frozen blocks~\cite{martinelli2004lectures} (i.e., a sufficiently large block has a unique steady state if one runs its dynamics freezing the boundaries): the locality of $H$ follows from that of the moves. Suppose the stationary measure is $\propto e^{-H(x)}$. Detailed balance gives
$W_u(x)/W_{-u}(x+u) = e^{-[H(x+u)-H(x)]}$. Since the lhs is local, so is the rhs: every $H$-difference between pairs of connected configurations is local. If all locally different pairs of configurations were connected by local moves, then this would immediately imply the locality of $H$. This last implication follows if we make one of two assumptions. If we assume that the model consists of single-site updates, conditioned on some finite window around the site, such that the rate is strictly positive for any state on the window, then any pair of configurations is connected by local moves and $H$ is manifestly local. More generally, assume that the chain of moves supported inside $A$ is irreducible for every fixed exterior. If $B$ shields $A$ from $C$, the frozen generator on $A$ depends only on $b$. Denote its stationary state as $\mu_b(A)$. Detailed balance means the generator of the dynamics is frustration-free, so the global steady state is also the steady state of the frozen chain, so the conditional distribution $P(A|b,c)$ is also a steady state for each $c$; uniqueness then gives $P(A\mid b,c)=\mu_b(A)$, and hence $I(A:C\mid B)=0$.  
Thus, there is a much tighter relation between detailed balance and Hammersley-Clifford in this setting: under this irreducibility assumption, a full-support local detailed-balance process must stabilize a local Hamiltonian. 
%

If one tries to derive a continuum process as the limit of a jump process, the argument above fails because each continuum increment sums over a large number of jumps, so channel-by-channel detailed balance is lost, and the weaker notion that survives for the continuum increments is unable to guarantee a local parent Hamiltonian (as the counterexample shows).  

\paragraph{Hyperuniform states cannot be reached rapidly.}
A natural attempt to discretize the counterexample might run as follows. Consider a system with three states per site, $\{ -1, 0, 1 \}$, indexed by their charge $q$, subject to the following two baths: a non-conserving zero-temperature bath that favors the state $0$, and drives both other states to it; and a finite-temperature bath that conserves $Q = \sum_i q_i$, and allows nearest-neighbor moves such as $0, \pm 1 \leftrightarrow \pm 1, 0$ and $00 \leftrightarrow +1, -1$. This fails to generate a hyperuniform steady state: processes like $0,0 \to 1,-1 \to 0, -1$ generate shot-noise fluctuations of the total charge, so the structure factor at zero momentum (which measures $\mathrm{var}(Q)$) is nonzero. 

The same obstruction applies whenever local charge-changing activity is nonzero and the local drift has summable correlations. Suppose that the jump process includes some charge-changing terms $r_e(x)$, each of which changes the charge by $\Delta Q_e$ (which in the above example is $\pm1$). Define the signed drift $b_Q(x)=\sum_e r_e(x)\Delta Q_e$ and the noise activity $v_Q(x)=\sum_e r_e(x)(\Delta Q_e)^2$. For a steady state with summable correlations of the local drift observables, $\mathrm{var}(b_Q)\le CN$ for some constant $C$. In the steady state, $\langle b_Q\rangle=0$ and $\langle Q^2\rangle$ is invariant. There are two contributions to its evolution, coming from the deterministic drift and the accumulation of variance from random increments. Balancing these gives $-2\operatorname{Cov}(Q,b_Q)=\langle v_Q\rangle\equiv Nr_2$. By Cauchy--Schwarz, $Nr_2\le2\sqrt{\mathrm{var}(Q)\mathrm{var}(b_Q)}$, and assembling these observations gives us the bound
\begin{equation}\label{quietness}
\frac{\mathrm{var}(Q)}{N} \geq \frac{r_2^2}{4C} > 0,
\end{equation}
so shot noise blocks hyperuniformity under these assumptions when $r_2>0$. The shot-noise floor is set by the minimum charge increment $\Delta Q$, and one recovers the counterexample in the previous section for a suitable $\Delta Q \to 0$ limit.

\paragraph{Is rapid mixing compatible with long-range CMI?}
Under the assumptions above, the argument ruled out the possibility of getting a long-range CMI through hyperuniformity. One might wonder whether there are other means to achieve a long-range CMI in the unique steady state of a jump process that mixes rapidly (as we were able to with a rapidly relaxing continuous process). 
We conjecture that this is impossible, based on the following heuristic argument. Since the process mixes rapidly and its steady state is unique, the steady state cannot have long-range order. All known examples where long-range conditional correlations persist in the absence of long-range order have some form of ``string order,'' i.e., there is a product of observables supported in $ABC$ that is ordered, though each individual factor is random. String order allows $A$ and $C$ to be correlated conditional on the state of $B$, even at early times (the string is always extended so it is not subject to light cones). 
However, in the steady state of a system with asynchronous stochastic jumps, the string order parameter in region $B$ will be flipped by single events in $B$ unless it is protected by some global symmetry of the dynamics (as it is for spontaneous strong-to-weak symmetry breaking \cite{Wang2026SWSSB}), or by some other structure such as synchronous updates designed to ensure that these local events cancel.
So any string decorrelates in its bulk, and long-range CMI cannot be achieved through this means. 
This step of the heuristic is in fact a theorem, proved in App.~\ref{app:string}: for a one-dimensional local jump process with a translation-invariant, spatially ergodic stationary law, persistent string order (with arbitrary bounded endpoint decorations) forces the string-flipping activity to vanish exactly; with full local support this is an exact conservation law of the dynamics, contradicting uniqueness of the stationary state on periodic chains. 
The remaining gap between this statement and the main conjecture is the first step of the heuristic --- that long-range CMI at short times must arise through a string. App.~\ref{app:hmm} closes this gap for binary steady states with an exact finite-state hidden-Markov representation, a strictly positive hidden transition matrix, and an active odd-flip rule, where exponential decay of the CMI itself can be proved; beyond that class, we are unable to rule out currently unknown mechanisms that avoid the ``string order'' route.

\section{Discussion}
\label{sec:discussion}

In terms of entropy production, the discrete and continuous classical worlds are surprisingly different: in the former, assuming full support and irreducible frozen blocks, a local process can only be in detailed balance with respect to a local Hamiltonian; in the latter, this restriction is gone, and the entropy production mandated by locality is ``hidden'' from the standard definitions of stochastic thermodynamics. The two worlds also differ in the nonequilibrium steady states they admit: in the discrete world, long-range CMI seems unachievable in the unique steady state of any rapidly mixing process, while the continuous world admits many straightforward examples with rapid drift relaxation. Interestingly, in this respect the discrete \emph{quantum} world looks more like the continuous classical one, admitting modes with no noise floor and rapid-mixing processes that generate long-range CMI \cite{NegariZouHsieh2026}. (Interestingly, our thermal implementation of this quantum example also requires infinite entropy production to sustain its steady state; a thermally regularized version pays entropy logarithmically in the Markov length, $\hat\sigma \simeq 2p(1-p)\ln\xi$ --- see App.~\ref{app:nzh}.)

Finally, we comment briefly on interacting mechanisms that evade the trade-offs of Sec.~\ref{sec:embedding}.  
The linear noise floor from the cold bath can be circumvented by tightly coupling the system to a nonequilibrium ratchet, which in turn is coupled to heat baths~\cite{StahlLakeKhemani2026}, such that every ``model A'' damping event coincides with a heat quantum $\Delta E$ transported from a hot to a cold bath, as in autonomous Maxwell demons \cite{Hill1977,StrasbergEtAl2013}. 
In this setting, wrong-way transitions are suppressed by $e^{-\Acal}$ with $\Acal = \Delta E(1/T_c - 1/T_h) \le E_{\rm loc}/T_c$ --- an effective temperature exponentially below the coldest reservoir, costing $\Acal$ per corrective event: effectively, this mechanism replaces the scaling $T_c \to \exp(-1/T_c)$, allowing for an exponentially larger $\xi$ at fixed $T_c$. 
%
However, all such strategies we have found require discrete events, so they cannot increase $\xi$ past the shot-noise floor of Eq.~\eqref{quietness}. Thus the general question of minimizing $\sigma$ at fixed $\xi$ for interacting systems remains open.


Our results raise many natural questions for future work. The most immediate one is to settle the conjecture on achieving long-range CMI in discrete jump processes (and to understand entropy production in such processes, if they exist). Other natural questions include generalizing our results to find a relation between quantum CMI and KMS detailed balance (cf.\ the quantum Gibbs sampler literature \cite{CKG2023}); identifying steady-state phase transitions at which the Markov length diverges, and understanding their thermodynamic signatures; extending our entropy-production floors beyond Gaussian theories to generic interacting field theories; and more generally incorporating our spatial locality restrictions (and Hammersley-Clifford) into the framework of stochastic thermodynamics.

\section*{Acknowledgements}
I used Claude Fable 5 and ChatGPT Astra 6 to develop the calculations and proofs in this work. I thank Yu-Hsueh Chen, Matthew Fisher, Sam Garratt, Tarun Grover, Hyunsoo Ha, Jake Hauser, Tim Hsieh, David Huse, J\'erôme Lloyd, Nick O'Dea, Shengqi Sang, Romain Vasseur, Cenke Xu, Carolyn Zhang, and Frank Zhang for helpful discussions.

\appendix

\section{CMI of the counterexample}
\label{app:cmi}

This appendix computes the conditional mutual information plotted in
Fig.~\ref{fig:cmi}.  The state is the regulated counterexample
on a ring of $N$ sites, $\Ccal(k) = T_1 + 4\sin^2(k/2)$; equivalently,
\begin{equation}
  \phi_i = E_i - E_{i-1} + \sqrt{T_1}\,u_i,
\end{equation}
with $E_i$ and $u_i$ independent standard Gaussians --- heights with
unit prior variance, plus a white regulator. 
We take $B$ to consist of two segments (or an annulus) of width $w$. 

For jointly Gaussian variables,
$I(A{:}C\mid B) = -\tfrac12\sum_i \ln(1-\rho_i^2)$, where $\rho_i$ are
the ``canonical correlations,'' i.e., the singular values of the matrix $R \equiv \Sigma_{AA}^{-1/2} \Sigma_{AC} \Sigma_{CC}^{-1/2}$, and the $\Sigma$'s are conditional covariance matrices.
For the equal-width buffers in App.~\ref{app:cmi}, both canonical
correlations have simple exact formulas, including at nonzero regulator.
Write $a=L_A$, $c=L_C$, $N=a+c+2w$, and define
\begin{equation}
  \kappa=2\operatorname{arsinh}(\sqrt{T_1}/2),\qquad
  u=\frac{(a-1)\kappa}{2},\quad
  v=\frac{(c-1)\kappa}{2},\quad
  t=(w+1)\kappa.
\end{equation}
For $a,c\ge2$, the two nonzero canonical correlations are
\begin{equation}
  \boxed{
  \rho_0^2=
  \frac{\cosh u\,\cosh v}
       {\cosh(u+t)\cosh(v+t)},
  \qquad
  \rho_1^2=
  \frac{\sinh u\,\sinh v}
       {\sinh(u+t)\sinh(v+t)}.
  }
\end{equation}
All remaining canonical correlations vanish, and
\begin{equation}
  I(A:C\mid B)
  =-\frac12\log(1-\rho_0^2)
   -\frac12\log(1-\rho_1^2) =\frac12\ln
 \frac{\sinh((L_A+2w+1)\kappa)\sinh((L_C+2w+1)\kappa)}
 {4\sinh^2((w+1)\kappa)\sinh^2(N\kappa/2)}.
 \label{eq:cmi2ch}
\end{equation}
In the small-regulator regime $N\kappa\ll1$,
\begin{equation}
 I(A{:}C\mid B)=\frac12\ln
 \frac{(L_A+2w+1)(L_C+2w+1)}{N^2(w+1)^2T_1}
 +O(T_1N^2).
\end{equation}

We expect a similar decomposition to hold for the annular geometry in $d = 2$: the sum over two channels is replaced by one over multipoles of the enclosed charge. We have not evaluated this in closed form: the plotted CMI in the main text is computed directly from covariance determinants.

\section{Keldysh analysis of response in the counterexample}
\label{app:keldysh}

In this appendix we show, using the Keldysh formalism, how the two implementations of the counterexample naturally give rise to distinct response functions, even though the processes are identical absent an external perturbation. We use the fact that coupling a system to an equilibrium bath at temperature $T$ and mobility $\Gamma$ adds the following term to the Keldysh action:

\begin{equation}
  \Delta S_a = \int_t\Big[\hat X\,\Gamma_a\,\frac{\delta H}{\delta X}
  \;-\; \Gamma_a T_a\,\hat X^2\Big],
  \label{eq:kbblock}
\end{equation}
The two readings of the counterexample have different Hamiltonians, and different system-bath couplings, so as to cancel out in the marginal process. The source is always added to $H$ as $H_0 + h X$.

\paragraph{Implementation A: long-range Hamiltonian, one bath.}
Coulomb Hamiltonian $\mathcal K_C(k) = 1/\hat k^2$, one
current-coupled (conserving) bath $\Gamma(k) = \Gamma_0\hat k^2$ at
temperature $T$.  Assembling one block:
\begin{equation}
  S^{A}_k[h] = \int_t\Big[\hat\phi\big(\partial_t +
  \Gamma_0\big)\phi \;-\; \Gamma_0 T\,\hat k^2\,\hat\phi^2
  \;-\; \Gamma_0\hat k^2\, h\,\hat\phi\Big];
  \label{eq:kbA}
\end{equation}
the conserving mobility cancels the Coulomb kernel in the drift (a
local current driven by the gradient of a long-range potential has a
local divergence), so $\Bcal_A = \Gamma_0$,
$\Dcal_A = \Gamma_0 T\hat k^2$, $S(k) = T\hat k^2$, and the single
bath is in detailed balance: $\sigma = 0$.

\paragraph{Implementation B: trivial Hamiltonian, two baths.}
$\mathcal K = 1$; a cold density-coupled bath $(\Gamma_c, T_c)$ and a
hot conserving bath $(\Gamma_h\hat k^2, T_h)$:
\begin{equation}
  S^{B}_k[h] = \int_t\Big[\hat\phi\big(\partial_t + \Gamma_c +
  \Gamma_h\hat k^2\big)\phi
  - \big(\Gamma_cT_c + \Gamma_hT_h\hat k^2\big)\hat\phi^2
  - \big(\Gamma_c + \Gamma_h\hat k^2\big)h\,\hat\phi\Big].
  \label{eq:kbB}
\end{equation}
In the limit $T_c \to 0$, $\Gamma_h \to 0$ with $\Gamma_hT_h$ fixed
(the cold block loses its noise, the hot block its friction), and
under the dictionary $\Gamma_c = \Gamma_0$,
$\Gamma_hT_h = \Gamma_0T$, the $h = 0$ parts of \eqref{eq:kbA} and
 \eqref{eq:kbB} become \emph{identical} term by term.  

\paragraph{Response.}
The source terms do not agree: implementation A carries
$-\Gamma_0\hat k^2\,h\hat\phi$, implementation B carries
$-\Gamma_0\,h\hat\phi$.  Hence
\begin{equation}
  \mathcal R_A(\omega,k) = \frac{\Gamma_0\hat k^2}{-i\omega+\Gamma_0},
  \qquad
  \mathcal R_B(\omega,k) = \frac{\Gamma_0}{-i\omega+\Gamma_0},
  \label{eq:kbchis}
\end{equation}
with identical $\Ccal(\omega,k)$.  

\section{Persistent string order forces a conserved charge}
\label{app:string}

This appendix proves the claim of Sec.~\ref{sec:jump} that in the
steady state of a local jump process, string order cannot survive
unless it is protected by an exact conservation law of the dynamics. (If a conservation law exists, it prevents rapid mixing.)
The strategy relies on translational invariance and spatial ergodicity, which is a consequence of the short-range correlations in the steady state. We also assume that the marginal distribution in the steady state has full support on the interaction range---i.e., for our two-site interaction window we assume that all four bit-strings occur with nonzero probability. Spatial ergodicity states that there are no long-range connected correlations in space. Translation-invariant, spatially ergodic distributions obey Birkhoff's theorem, which states that for any bounded local observable $g$, the space-average of any observable converges to its ensemble average. 
To keep the notation simple we will state the theorem for bits with two-site updates, and consider the parity string $S_n(x) = (-1)^{x_1 + x_2 + \ldots x_n}$, but the argument manifestly works for any finite-size alphabet and any string that is a product over sites. The theorem states:

\begin{theorem} (informal). \label{thm:string}
Under the assumptions above, $\langle S_n \rangle \to 0$ as $n \to \infty$ in the steady state.  
\end{theorem}

\begin{proof}
We sketch the proof. Again, to keep the combinatorics transparent we assume that all updates act on nearest-neighbor pairs of sites. Define $r_i(x_i, x_{i+1})$ as the probability that an update acting on that configuration will flip the value of $S_n$ (for $i, i+1$ inside the string), and define $a = \langle r_i \rangle_{\mathrm{s.s.}}$ where the average is over the steady-state probability distribution (and is independent of $i$ by assumption). The Markov process generator $\mathcal{L}$, acting on $S_n$, gives
\begin{equation}
\mathcal{L} S_n = - 2 S_n \sum_{i = 1}^{n-1} r_i + G_n, 
\end{equation}
where $G_n$ are boundary terms of order unity that can be bounded by a constant $G$. Since $\langle \mathcal{L} S_n \rangle_{\mathrm{s.s.}} = 0$ by stationarity [we will drop the subscript from now on], we can bound
\begin{equation}\label{lsn}
\left|\left\langle S_n \left(n^{-1} \sum_{i=1}^{n-1} r_i \right) \right\rangle\right| \leq G/n.
\end{equation}
Define $\overline{r}_n \equiv n^{-1} \sum_{i=1}^{n-1} r_i$. Since the magnitude of $S_n$ is at most $1$, we can also bound 
\begin{equation}
|\langle S_n \overline{r}_n \rangle - a \langle S_n \rangle| \leq |\overline{r}_n - a |.
\end{equation}
By Birkhoff's theorem the latter quantity goes to zero, so for large $n$, $a^{-1} \langle S_n \rangle \to |\langle S_n \overline{r}_n \rangle| \to 0$ by Eq.~\eqref{lsn}. So either $a = 0$ or $\langle S_n \rangle = 0$. Since $a$ is an average over non-negative rates, and the steady state has support on all configurations, if $a = 0$ on average the rate at which parity flips must be identically zero. If this is true then parity is conserved and the dynamics is non-ergodic, contradicting our assumptions. 
\end{proof}
The value of this proof, and the reason it improves on the argument in the main text, is that we made no assumptions about the correlation between the value of the string and the local rates in the region it covers. 

\section{Short-range CMI for steady states with hidden Markov structure}
\label{app:hmm}

This appendix states and proves the one-dimensional obstruction
announced in Sec.~\ref{sec:jump}:

\begin{theorem}[finite-state case]\label{thm:hmm}
Let $\pi$ be a translation-invariant stationary law on
$\{\pm1\}^{\mathbb Z}$ of a finite-range jump process with bounded
rates and at least one nonzero bulk rule flipping an odd number of
spins, and suppose $\pi$ has an exact edge-emitting hidden-Markov
representation with $D$ hidden states and strictly positive hidden
transition matrix, $T_{ab} \ge \delta > 0$.  Then the two-sided CMI
across a buffer of width $\ell$ obeys
$\Ccal_\ell \le C e^{-c\ell}$, with $(C,c)$ depending only on
$(D,\delta)$ for deterministic readout --- and only on
$(D,\delta,\eta)$ for stochastic readout whose nonzero labeled
transition probabilities are $\ge \eta$ --- uniformly in the jump
rates, their range, and the entropy production.
\end{theorem}

\paragraph{Two-state hidden layer.} First we consider a much simpler version of the general problem. We take a system with hidden bits $Y_i$ on every edge, visible spins $X_i$ on every vertex $i$ (right of edge $i$), and a deterministic function $X_i = f(Y_i, Y_{i+1})$ coupling them (this is called ``deterministic readout''). The underlying probability distribution is over $Y_i$, which obey a Markov chain in the spatial direction, but $X_i$ ``reads'' this Markov chain through the deterministic function. In this simple case there are only sixteen functions that could exist, and they fall into four classes: (i)~two constant functions, which lead to product distributions; (ii)~four cases where $X_i = \pm Y_i, \pm Y_{i+1}$, where $X$ inherits the Markov property; (iii)~eight functions for which at least one output $X_i$ reveals $Y_i$ and $Y_{i+1}$, so (by the Markov property) conditioning on $X_i$ decouples the left and right states on $Y$ and thus (by data processing) on $X$, so the CMI vanishes; and (iv)~two functions $X_i = \pm Y_{i} Y_{i+1}$. These parity functions are the only case that might admit long-range CMI. The mechanism for long-range CMI is, however, precisely string order: $\langle Y_i Y_{i+n} \rangle = \langle \prod_{j \in [i, i+n -1]} X_j \rangle$. A zero-dimensional Markov chain with entirely positive rates certainly mixes, so the result in the previous section rules out this case, and (for a two-dimensional hidden state space) thus rules out long-range CMI. 

\paragraph{General case: strategy.}
Either visible observations erase hidden-state ambiguity, or a
hidden parity survives. In the first case, there are some configurations in which the hidden variables are revealed to the visible ones, and this decorrelates the hidden Markov chain and cuts off the CMI. This intuition is converted to an exponential bound by Travers's result~\cite{Travers2014}.
In the second case, either the hidden parity produces string order (in which case we rule it out as before) or it is completely hidden, in which case it does not affect any properties of the visible layer and thus does not contribute CMI.

\paragraph{Predictive memory.}
Set
\[
  h_n=H(X_n\mid X_1^{n-1}),\qquad h=\lim_n h_n.
\]
The entropy chain rule gives
\[
  \mathcal C_\ell(X)=\sum_{n>\ell}(h_n-h).
\]
Thus exponential convergence of $h_n$ suffices. All these quantities
are finite: past and future communicate through a hidden state,
so their mutual information is at most $\log D$.

\paragraph{The binary support dichotomy.}
Two hidden states are mergeable if they can emit the same word and
reach the same hidden state. They are immediately incompatible if
they cannot emit the same first symbol. Either every pair merges
within two steps or is immediately incompatible, or
\[
  X_i=\sigma\,s(S_{i-1})s(S_i)
\]
for a nonconstant hidden sign $s$.

Indeed, take a pair $a,b$ violating the first alternative.
Failure to merge in one step, together with $T>0$ and the binary
alphabet, forces their labels to every destination $j$ to be
deterministic and opposite, say $f(j)$ and $-f(j)$.
Immediate compatibility makes $f$ nonconstant. Failure to merge
in two steps then forces every row indexed by $f=+1$ to be
opposite every row indexed by $f=-1$. All positive rows therefore
equal some row $g$, and all negative rows equal $-g$.
Comparing with row $a$ gives $g=f(a)f$, hence
\[
  f(i,j)=f(a)f(i)f(j),
\]
the claimed parity form.

In the nonexceptional case, Travers's theorem gives exponential
convergence of $h_n$
\cite[Theorem~2 and Remark~3]{Travers2014}.
For uniformity, his finite-block coupling uses a block length
bounded by some $K(D)$: there are only finitely many labeled
support patterns. Positive block entries are at least
$\eta^{K(D)}$, and stationary hidden weights are at least $\delta$.
His explicit estimates therefore give uniform constants.

\paragraph{An asymmetric parity sector produces string order.}
Absorb $\sigma$ into the visible spins and write
\[
  Y_i=s(S_i),\qquad X_i=Y_{i-1}Y_i.
\]
If the law of $Y$ is not invariant under $Y\mapsto-Y$, some bounded
odd function $G$ of a finite block has mean $g\ne0$. The functions
$Y_0G(Y_{-q}^{0})$ and $Y_nG(Y_n^{n+q})$ are sign-even, so they
depend only on the corresponding visible endpoint blocks.
Consequently there are bounded decorations $f_L,f_R$ such that
\[
  O_n
  =f_L(X_{-q+1}^{0})
    \left(\prod_{j=1}^{n}X_j\right)
    f_R(X_{n+1}^{n+q})
  =G(Y_{-q}^{0})G(Y_n^{n+q}).
\]
Mixing of the positive hidden chain gives
$\mathbb E O_n\to g^2$.

Let the update window have width $R$, including every spin read
or flipped, and let $r_i$ be the total rate of odd-spin flips in
its translate beginning at $i$. Write $a=\mathbb E r_i$.
Interior odd flips negate $O_n$, while only finitely many windows
touch its endpoints. Stationarity therefore gives
\[
  0=\mathbb E[LO_n]
   =-2\mathbb E\left[
        O_n\sum_{i=1}^{n-R+1}r_i
      \right]+O(1).
\]
The visible process is spatially ergodic, so
$n^{-1}\sum_i r_i\to a$ in $L^1$. Since $O_n$ is uniformly bounded,
division by $n$ gives $ag^2=0$. But every finite visible word has
positive probability in the parity representation: choose an
initial sign, propagate it using the visible word, and choose
hidden states in the required sign classes. Thus the nonzero
odd rule implies $a>0$, a contradiction. The law of $Y$ must be
sign-symmetric.

\paragraph{An invisible symmetry does not affect anything.}
The only case that remains is the trivial one: the hidden layer has an exact symmetry that nothing in the visible layer ever detects (a ``strong symmetry'' of the spatial evolution). In this case it produces no structure in the visible layer and can be ignored.

\section{Entropy production of a quantum steady state with long-range CMI}
\label{app:nzh}

\paragraph{Model.}
Write $c_j=(a_j+ib_j)/2$ and $d_j=(a_j+ib_{j+1})/2$, with
$a_j,b_j$ Majoranas. The open-chain fermionic model of
Ref.~\cite{NegariZouHsieh2026}, with boundary damping completed, has jumps
\begin{equation}
 \sqrt{1-p}\,c_j\ (1\le j\le N),\qquad
 \sqrt p\,d_j\ (1\le j<N),\qquad
 \frac{\sqrt p}{2}b_1,\quad\frac{\sqrt p}{2}a_N,
 \label{eq:nzhopen}
\end{equation}
where $0<p<1$ and $\mathcal D[L]\rho=L\rho L^\dagger-
\{L^\dagger L,\rho\}/2$.
For a local spin realization, introduce auxiliary fermions
$g_j=(\gamma_j+i\delta_j)/2$, replace each jump $F$ anchored at $j$
by $i\gamma_jF$, and add unit-rate jumps
$-i\gamma_j\delta_j$ ($j\le N$), $i\gamma_j\gamma_{j+1}$ ($j<N$),
and $\gamma_1$.
Jordan--Wigner ordering $g_1,c_1,\ldots,g_N,c_N$ gives jumps on at most
four consecutive spins; $\gamma_1$ is the leftmost Pauli $X$ and removes
the parity degeneracy. The presence of boundary terms ensures a unique steady state in the spin representation; rapid mixing and long-range CMI were already established. The origin of the long-range CMI is that the Lindbladian contains no terms repopulating the $k = 0$ mode of the $c$ fermions. Indeed, in steady state, $\langle c^\dagger_k c_k \rangle \sim k^2$, so the steady state is hyperuniform, analogous to our classical example.

\paragraph{Implementation.}
We implement the system-bath coupling in the ``collision model'' framework: system fermionic operators are alternately coupled to ancillas coupling respectively to the $c$ and $d$ fermions. Each ancilla has Hamiltonian $\varepsilon \tau^z$, and is initialized at temperature $T$, i.e., excited-state probability
$f=(e^{\varepsilon/T}+1)^{-1}$. The ancilla couples to the system via 
$V(t)=g(t)(F\otimes s^++F^\dagger\otimes s^-)$, where
$s^+=|1\rangle\langle0|$ raises the ancilla. 
The time-dependent drive is needed to allow jump operators to try to ``cool'' fermions in distinct bases.
In the weak-coupling limit the baths have the following jump operators:
\begin{equation}
 \sqrt{(1-p)(1-f)}\,c_j,\quad\sqrt{(1-p)f}\,c_j^\dagger,
 \qquad\sqrt{p(1-f)}\,d_j,\quad\sqrt{pf}\,d_j^\dagger.
 \label{eq:nzhthermal}
\end{equation}
At $f=0$, both channels empty the $k=0$ mode: pumping vanishes there,
but loss remains finite. For $f \neq 0$ the Markov length scales as $\xi \sim \sqrt{p(1-p)/f}$. Given this data and the steady-state occupation factors, the energy current into each bath is just the change in its occupation, and the entropy production
per site and relaxation time is
\begin{equation}
 \hat\sigma=\frac{\varepsilon}{T}\int_{-\pi}^{\pi}
 \frac{dk}{2\pi}(J_c+J_d)
 =p(1-p)(1-2f)\log\frac{1-f}{f}
 =2p(1-p)\log\xi+O(1).
 \label{eq:nzhtrade}
\end{equation}
This argument was for an explicit bath but it manifestly applies to any bath that cools the $k = 0$ mode to vacuum at unit rate: the combination of rapid dissipation and occupation near zero forces the cold bath to rapidly absorb heat, leading to a diverging entropy production rate in general.

\bibliography{refs}

\end{document}